\documentclass[11pt,aps,pra,reprint]{revtex4-1}
\usepackage{amsmath}    
\usepackage{graphicx}   
\usepackage{bm}

\begin{document}
\title{Energy-dependent frame transformation theory for dissociative recombination}
\author{D\'{a}vid Hvizdo\v{s}}
\affiliation{J. Heyrovsk\'{y} Institute of Physical Chemistry, ASCR,
Dolej\v{s}kova 3, 18223 Prague, Czech Republic}
\affiliation{Institute of Theoretical Physics, Faculty of Mathematics and Physics, Charles University in Prague, V Hole\v{s}vi\v{c}k\'{a}ch 2, 180 00 Prague, Czech Republic}
\author{Chris H.~Greene}
\affiliation{Department of Physics and Astronomy, Purdue University, West Lafayette,
Indiana 47907, USA}
\affiliation{Purdue Quantum Science and Engineering Institute, Purdue University, West Lafayette, Indiana 47907, USA}
\author{Roman \v{C}ur\'{\i}k} \email{roman.curik@jh-inst.cas.cz}
\affiliation{J. Heyrovsk\'{y} Institute of Physical Chemistry, ASCR,
Dolej\v{s}kova 3, 18223 Prague, Czech Republic}
\date{\today}

\begin{abstract}
The energy-dependent frame transformation theory of Gao and Greene 1990 [Phys. Rev. A {\bf 42}, 6946 (1990)]
is extended to yield quantitatively accurate description of the
dissociative recombination process. Evidence is presented to show that direct application of the original theory 
leads to inaccurate cross sections. A major revision, based on an
interaction-free back-propagation of the Born-Oppenheimer solutions, markedly improves the frame transformation
theory, reducing its average error by orders of magnitude.
The original theory and its extension are tested on the previously explored 2D model that is 
tailored to describe the singlet ungerade states
of molecular hydrogen. The 2D model can be solved exactly (within the numerical accuracy) without implementing the 
Born-Oppenheimer approximation. These exact results then serve as a benchmark for the frame 
transformation theory developed in this paper.
\end{abstract}
\maketitle

\section{\label{sec-intro}Introduction}

Rovibrational frame transformation (FT) theory \cite{Chang_Fano_1972} 
divides the electronic space into two parts: the inner (body frame) and the outer 
(laboratory frame) regions. Electronic and nuclear
Hamiltonians are considered decoupled in the outer region and hence the solutions
of the outer region Schr\"{o}dinger equation are linear combinations of products 
of the electronic and rovibrational wave functions.
Solutions of the system in the inner region are assumed to be quasi-separable 
Born-Oppenheimer wave functions. The role of the frame transformation theory is to 
smoothly connect the independent solutions in these two regions.

In treatments of inelastic collisions
between electrons and neutral molecules the adiabatic-nuclei approximation \cite{Chase_1956} 
has often been assumed
\cite{Morrison_greenbook1,Cascella_CGS_2001,Curik_FAG_cyc2_2002}. Formally this technique
can be viewed as an application of the FT procedure carried out at infinite electronic radius.
Therefore, the FT theory should not
be confused with the adiabatic-nuclei approximation, as the former exploits the Born-Oppenheimer
approximation (BOA) only at small electronic distances, while the latter employs it over the entire
electronic space.

Combinations of the FT approach with multichannel quantum defect theory (MQDT)
\cite{Seaton_RPP_1983,Orange_review} have been previously applied to treat numerous electron-cation
collision and molecular photofragmentation systems (see, e.g., Refs.
\cite{Jungen_Dill_1980,Jungen_PRL_1984,Greene_Jungen_AAMP_1985,Hamilton_Greene_PRL_2002,Takagi_HeH_2004,
Curik_Greene_PRL_2007,Curik_Greene_JCP_2017,Ayouz_Kokoouline_Atoms_2016,Khamesian_AK_Atoms_2018,Kokoouline_Greene_PRA_2003}) and
still more applications can be found in Rydberg spectroscopy 
\cite{Orange_review}. The cornerstone of these studies is the body-frame
quantum defect (or phase shift) $\mu(R,\epsilon)$ that describes a phase
gained by the scattered (or Rydberg) electron inside the molecular core. 
The difficulty in application of the FT theory has always been in the choice of the body-frame energy $\epsilon$ at
which the quantum defect $\mu(R,\epsilon)$ is determined. This crucial question needs to be addressed in order 
to calculate vibrational matrix elements of operators such as $\sin{\pi \mu(R,\epsilon)}$ and 
$\cos{\pi \mu(R,\epsilon)}$.
This problem does not arise in cases where the energy dependence of $\mu$ can be neglected, as in the vast
majority of the MQDT studies carried out up to this date. However, 
two different theoretical approaches were developed to account for the energy dependence of the inner solutions.
A common element of these two treatments is the introduction of the electron-molecule compound potential-energy
curves, along which the nuclei move when the scattered electron is inside the inner region. Unlike the
bound-state problems, where the BOA potential-energy curves are well defined, for the continuum electronic
energies there is not an obvious {\it a priori} way to connect electronic BOA energies 
at nuclear coordinates $R$ with those determined at $R+\delta$. This has led
to development of the two different FT theories that differ in their choices of the BOA potential-energy curves.

In the first approach \cite{Greene_Jungen_PRL_1985} (further extended in 
Refs.~\cite{Gao_Greene_JCP_1989,Robicheaux_1991}), the potential-energy curves
in the continuum were chosen such that the quantum defect $\mu(R,\epsilon)$ does not depend on the 
internuclear distance $R$.  The difficulty of this method lies in finding these potential-energy curves.
Once they are known, the frame
transformation matrix can be reduced, in this case, only to a Franck-Condon overlap integral between
the vibrational states of the target and those of the compound \cite{Greene_Jungen_PRL_1985}.

In the second approach \cite{Gao_Greene_PRAR_1990}, the compound BOA potential-energy curves are
chosen explicitly as curves parallel to the curve of the target molecular system. The vertical distance
of the compound curves from the target curve correlates with the collision energy of the electron in the incident
vibrational channel.
The method was successfully tested for resonant electron-impact vibrational 
excitation of N$_2$ and for determination of vibrational levels of the H$_2$ $B''$ state. However,
there has been no application to dissociative recombination.

In the present study we adopt the second approach to study indirect dissociative recombination with
energy-dependent frame transformation theory. Part of our motivation for this choice lies in the 
similarity between
the compound bound potential energy curves (Rydberg curves) and the target cation curve. In order
to assess the results, the approximate FT theory will be applied to a model 2D system
\cite{Hvizdos_VHGRMC_PRA_2018} tailored to describe dissociative recombination of H$_2^+$ through
its singlet ungerade channels. This 2D model can be solved exactly (within numerical accuracy)
\cite{Hvizdos_VHGRMC_PRA_2018,Curik_HG_PRA_2018}, which bypasses all physical approximations and
thus serves as an exact benchmark for the approximate FT theory.

\section{\label{sec-GGFT}Energy-dependent frame transformation}

The goal of the frame transformation theory is to obtain the scattering (or reactance) matrix 
describing the coupling of asymptotic channels at all distances beyond some fixed electronic 
radius $r_0$. The distance $r_0$ is chosen as small as possible to ensure validity of the Born-Oppenheimer
approximation inside the electronic volume confined by $r_0$.
For the present model Hamiltonian $H$ which has no long-range power-law potential coupling terms, 
the chosen value $r_0$ is also such that
for $r \ge r_0$ all the interaction terms in $H$ apart from the Coulomb potential are negligible.
In this section we briefly
summarize the basic steps of the energy-dependent FT of Ref.~\cite{Gao_Greene_PRAR_1990} since this
theory is the starting point of the present study.

A set of linearly-independent Born-Oppenheimer solutions can be written in the inner region for $r<r_0$ as
\begin{equation}
\label{eq-gg-psi}
\psi_{i'}(R,r) = \phi_{i'}(R)\, F_{i'}(R; r)\;,
\end{equation}
where $\phi_{i'}(R)$ are vibrational eigensolutions of the target (or compound) nuclear Hamiltonian with
eigenenergies $E_{i'}$. The electronic solutions $F_{i'}(R; r)$ are normalized electronic
BOA eigensolutions at fixed
coordinate $R$ with eigenenergies $\epsilon_{i'}$. The total energy is $E = E_{i'} + \epsilon_{i'}$.

As is standard in quantum defect treatments, for $r\ge r_0$ the interaction with the molecular core is assumed to be
solely the Coulomb potential, whereby we can write the electronic inner-region Born-Oppenheimer solutions at $r_0$ as
\begin{alignat}{1}
\nonumber
F_{i'}(R; r) =& \,N(R,\epsilon_{i'})\, \left[ f_{\epsilon_{i'}}(r_0) \cos\pi\mu(R,\epsilon_{i'})\right. \\
\label{eq-gg-F}
-& \left. g_{\epsilon_{i'}}(r_0) \sin\pi\mu(R,\epsilon_{i'})\right].
\end{alignat}
Here $f_{\epsilon_{i'}}(r_0) \equiv f_{i'}(r_0)$ and $g_{\epsilon_{i'}}(r_0) \equiv g_{i'}(r_0)$ 
are regular and irregular Coulomb functions evaluated for the energy
$\epsilon_{i'}$. The normalization factor $N(R,\epsilon_{i'})$ was introduced to ensure the volume normalization
of $F_{i'}(R; r)$, since the term in brackets is just a surface term not possessing any kind of normalization.
It has been shown previously
\cite{Lee_PRA_1974,Gao_Greene_JCP_1989}
that the normalization factor can be evaluated solely from the surface properties as
\begin{equation}
\label{eq-gg-norm}
N(R,\epsilon_{i'}) = \left[ \frac{\partial\mu(R,\epsilon_{i'})}{\partial \epsilon}
+ \frac{1}{2}W(R,\epsilon_{i'}) \right]^{-1/2},
\end{equation}
with
\begin{alignat}{1}
\nonumber
W(R,\epsilon) =& \left( [ f_\epsilon,g'_\epsilon] + [g_\epsilon,f'_\epsilon] \right)
\sin\pi\mu(R,\epsilon) \cos\pi\mu(R,\epsilon) \\
\label{eq-gg-w}
-& [f_\epsilon,f'_\epsilon]
\cos^2\pi\mu(R,\epsilon) - [g_\epsilon,g'_\epsilon] \sin^2\pi\mu(R,\epsilon)\, ,
\end{alignat}
where $[f,g]$ denotes the Wronskian of functions $f$ and $g$, and $f'\equiv \partial f/ \partial \epsilon$.

In the outer region ($r>r_0)$ the independent solutions (\ref{eq-gg-psi}) can be 
written as a linear combination of channel functions (close-coupling expansion)
\begin{equation}
\label{eq-gg-asym}
\psi_{i'}(R,r) = \sum_i \phi_{i}(R) \left[ f_{i}(r) I_{i i'} - g_{i}(r) J_{i i'} \right]
\end{equation}
Matching of equations (\ref{eq-gg-psi}) and (\ref{eq-gg-asym})  at $r_0$ determines the matrices
\begin{eqnarray}
\nonumber
I_{i i'} &=& [f_{i'},g_{i}] C_{i i'} - [g_{i'},g_{i}] S_{i i'},\\
\label{eq-gg-IJ}
J_{i i'} &=& [f_{i'},f_{i}] C_{i i'} - [g_{i'},f_{i}] S_{i i'},
\end{eqnarray}
with
\begin{eqnarray}
\nonumber
C_{i i'} = \int dR\, \phi_i(R) N(R,\epsilon_{i'}) \cos\pi\mu(R,\epsilon_{i'}) \phi_{i'}(R)\;,\\
\label{eq-gg-SC}
S_{i i'} = \int dR\, \phi_i(R) N(R,\epsilon_{i'}) \sin\pi\mu(R,\epsilon_{i'}) \phi_{i'}(R)\;.
\end{eqnarray}
The short-range $K$-matrix is then obtained as $\underline{K} = \underline{J}\, \underline{I}^{-1}$.

\begin{figure}[htb]
\includegraphics[width=0.48\textwidth]{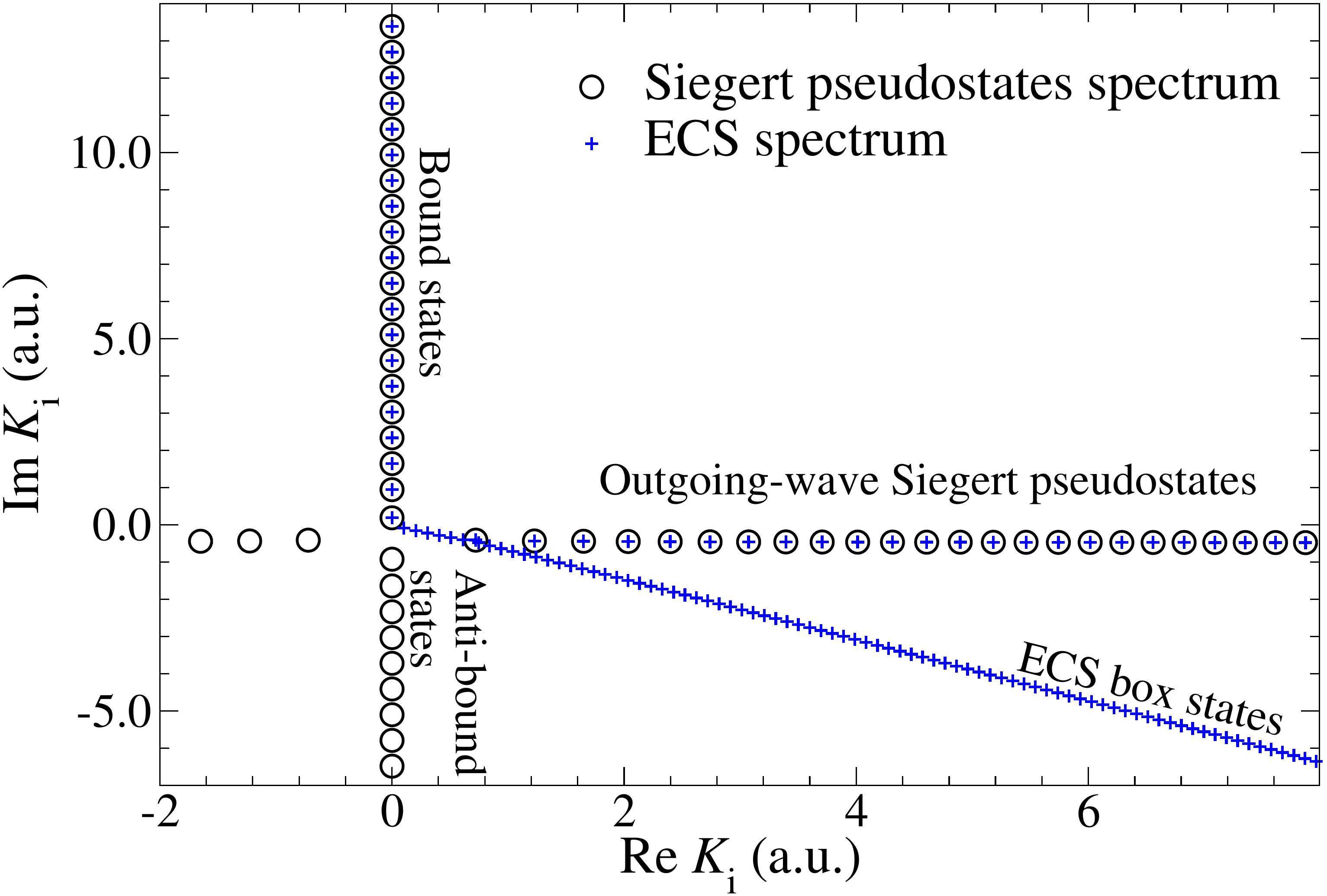}
\caption{\label{fig-kplane}
Distribution of the Siegert pseudostate poles (circles) defined by $R_0$~=~15 bohr and of the
ECS poles with the complex contour (\ref{eq-ECS-countour}) (crosses) in the complex nuclear momentum
plane.
}
\end{figure}

In a manner similar to our previous studies 
\cite{Hamilton_Greene_PRL_2002,Kokoouline_Greene_PRA_2003,Curik_Greene_MP_2007,Curik_Gianturco_2013,Curik_Greene_JCP_2017},
instead of real vibrational functions, we employ a complex
vibrational basis $\phi_{i}(R)$. Such a basis can be obtained by applying the Siegert boundary condition
\cite{Siegert_1939,Tolst_sieg_1997,Tolst_sieg_1998}
at the nuclear coordinate $R = R_0$ or, as is done in the present study, by solving the nuclear Schr\"{o}dinger 
equation along a contour $Z$ in the complex plane. The zero-value boundary condition is applied here at both ends
of the contour. We utilize the technique of exterior complex scaling (ECS)
\cite{Simon_ECS_PLA_1979,McCurdy_Martin_JPB_2004}
with the complex contour chosen as
\begin{equation}
\label{eq-ECS-countour}
Z = \left\{
\begin{array}{ll}
 R, & \mathrm{for}\; R \leq R_0, \\
 R_0 + e^{i\theta(R-R_0)}, & \mathrm{for}\; R_0 < R \leq R_m ,
\end{array}
\right.
\end{equation}
where $R$ is a real parameter along the complex contour $Z$, $R_0 = 15$ bohr denotes the bending point,
$\theta = 40^{\circ}$ is the bending angle, and $R_m = 40$ bohr parameterizes the final point $Z_m$ of the
complex contour. The spectra of the Siegert and ECS systems are similar, as is shown in Fig.~\ref{fig-kplane}.
The ECS spectrum contains a branch coinciding with the bound and outgoing-wave Siegert pseudostates, while it 
is missing the anti-bound and incoming-wave Siegert pseudostates branches. The linear branch close to
negative 40$^\circ$ corresponds to ECS states defined by the zero boundary condition at the end
of the bent interval. Note, that the coincidence of the outgoing-wave Siegert spectrum and
the ECS spectrum is due to the coincidence of the Siegert boundary $R_0$ with the bending point of
the ECS contour (\ref{eq-ECS-countour}).

The resulting cross section obtained with the selected subset of Siegert states \cite{Hvizdos_VHGRMC_PRA_2018} and 
those obtained with the ECS states are numerically equal except for some small energy windows, at which
high sensitivity to the completeness of the vibrational basis can be observed.
Our experience shows that the ECS states are complete to better numerical accuracy, when compared to the 
completeness of the bound and outgoing-wave subset of Siegert pseudostates.

Completeness of the vibrational basis also affects the symmetry of the resulting $K$-matrix. 
The symmetry of the $K$-matrix is also further disturbed by the right-index dependence of the
body-frame energy $\epsilon$ in Eqs.~(\ref{eq-gg-SC}).
We have observed that the symmetry of the short-range $K$-matrix is essential for
stability of the final cross sections. Therefore, in the present study the
$K$-matrix is artificially symmetrized by the replacement $K \rightarrow (K + K^T)/2$ before the closed channels 
are eliminated in the MQDT calculation.
A similar ad-hoc symmetrization step was also reported
as being necessary in a previous application of the energy-dependent FT theory to dissociative electron attachment
of electrons colliding with H$_2$ \cite{Robicheaux_1991}.
It is important to note that in the ECS (or Siegert pseudostates) basis the 
$K$-matrix should be symmetric but not Hermitian, owing to the missing complex conjugate of $\phi_i(R)$ in
Eq.~(\ref{eq-gg-SC}).

After the Cayley transformation of the short-range $K$-matrix to the short-range $S$-matrix
\begin{equation}
\underline{S} = \left( \underline{1} + i \underline{K} \right) 
                \left( \underline{1} - i \underline{K} \right)^{-1}\; ,
\end{equation}
the physical $S$-matrix is obtained by the standard closed-channel elimination technique of the MQDT:
\begin{equation}
\label{eq-gg-elimc}
\underline{S}^{\mathrm{phys}} = \underline{S}^{oo} - \underline{S}^{oc}\left[
\underline{S}^{cc}-e^{-2 i \underline{\beta}(E)} \right]^{-1} \underline{S}^{co} \;,
\end{equation}
where the superscripts $o$ and $c$ denote open and closed sub-blocks in the
short-range $S$-matrix, respectively.
The diagonal matrix $\underline{\beta}(E)$ describes effective Rydberg quantum
numbers with respect to the closed-channel thresholds $E_i$:
\begin{equation}
\label{eq-elimb}
\beta_{ij} = \frac{\pi}{\sqrt{2(E_i-E)}}\delta_{ij}\;.
\end{equation}

Finally, the utilization of the complex nuclear basis with the outgoing-wave boundary conditions
at $R_0$ allows us to compute the dissociative flux solely on the electronic surface. 
The defect of unitarity of the electronic physical $S$-matrix $\underline{S}^{\mathrm{phys}}$,
i.e. the missing electronic flux of the system, can be identified 
\cite{Hamilton_Greene_PRL_2002} with the dissociative flux
\begin{equation}
\label{eq-gg-sigma}
\sigma_{i'}(\epsilon_{i'}) = \frac{\pi}{2 \epsilon_{i'}} \left[ 1 - \sum_i S^{\mathrm{phys}\,\dag}_{i' i}
S^{\mathrm{phys}}_{i i'} \right]\;.
\end{equation}
The validity of this ansatz was previously confirmed using the 2D model \cite{Hvizdos_VHGRMC_PRA_2018}.
This approach does have one important limitation, in that it is unable to separate the partial DR cross sections 
in different dissociation channels, because the method computes only the total dissociative flux.

\section{\label{sec-2d}Application of the FT theory to the 2D model}

The 2D model, has two different modes of fragmentation associated with the competing
dissociation and ionization (or detachment) channels, and is
described by the Schr\"{o}dinger equation
\begin{equation}
\label{eq-2d-Schrodinger}
\left[ H_{\mathrm{n}}(R) + H_{\mathrm{e}}(r) + V(R,r) - E\right] \psi(R,r) = 0 \;,
\end{equation}
where
\begin{eqnarray}
\label{eq-2D-Hn}
H_{\mathrm{n}}(R) &=& - \frac{1}{2 M}\frac{\partial^2}{\partial R^2} + V_0(R)\;, \\
\label{eq-2D-He}
H_{\mathrm{e}}(r) &=& - \frac{1}{2}\frac{\partial^2}{\partial r^2} + 
\frac{l(l+1)}{2 r^2} - \frac{1}{r}\;.
\end{eqnarray}
The potential curve $V_0(R)$ describes the vibrational motion of the target molecule.
The potential energy function $V(R,r)$ couples the electronic and nuclear degrees of freedom and is set
to approximately describe the singlet ungerade Rydberg series of H$_2$ and the singlet ungerade low-energy
scattering of electrons by H$_2^+$. More details and the exact forms of the potentials
$V_0(R)$ and $V(R,r)$ can be found in 
Refs.~\cite{Hvizdos_VHGRMC_PRA_2018,Curik_HG_PRA_2018}.

\begin{figure}[bht]
\includegraphics[width=0.48\textwidth]{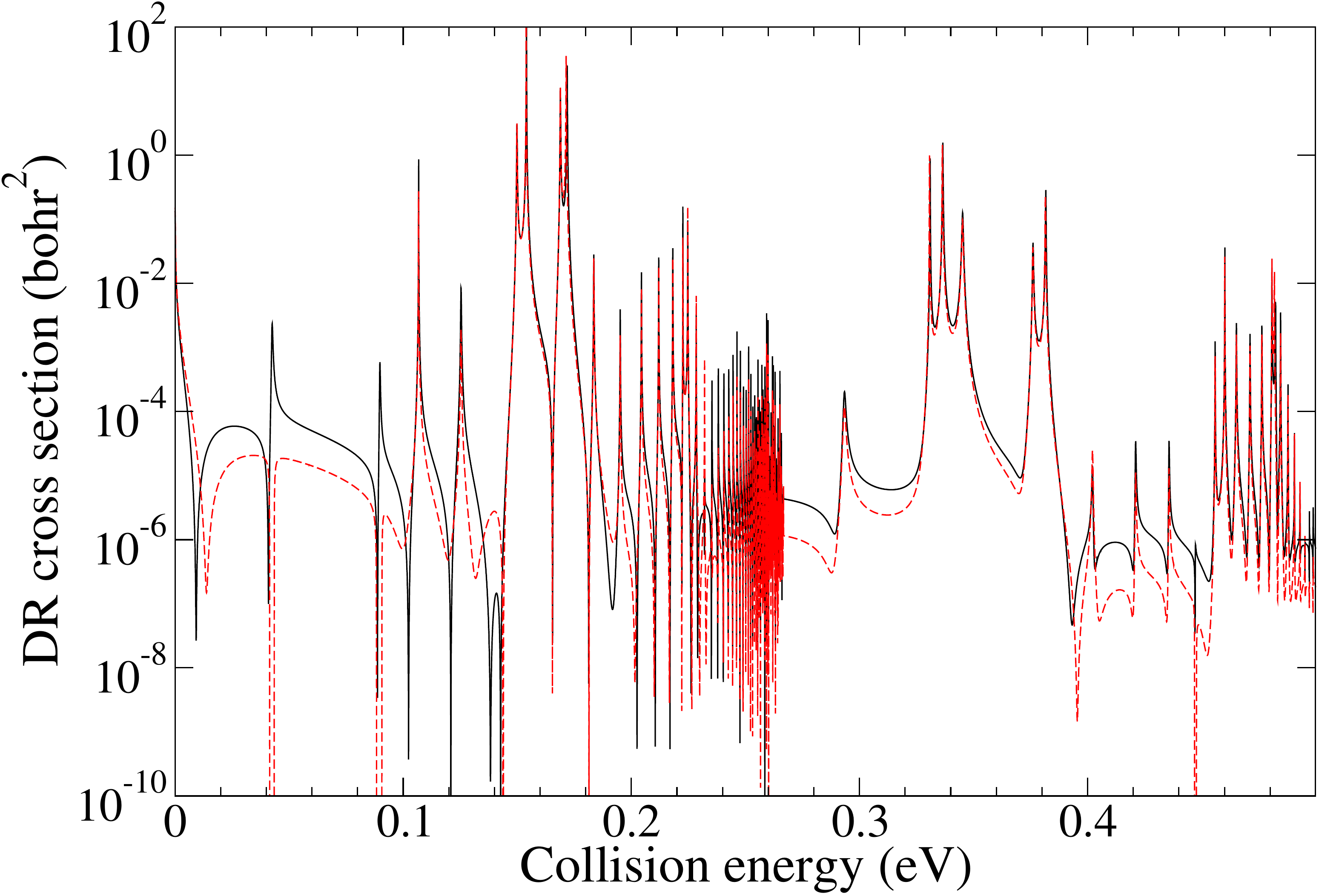}
\caption{\label{fig-gg_07}
DR cross sections obtained from the exact solution of the 2D Hamiltonian (full line) and
from the direct application of the present energy-dependent FT theory (dashed line).
}
\end{figure}

Direct application of the theory presented in the previous section leads to inaccurate resonance
lineshapes and cross section magnitudes, as is
shown in Fig.~\ref{fig-gg_07}. The essentially exact solution represented by the black curve was
obtained with the 2D $R$-matrix method \cite{Curik_HG_PRA_2018}, while the red dashed curve represents
results obtained with the energy-dependent FT of Ref.~\cite{Gao_Greene_PRAR_1990} in the form that was
summarized above and
extended to treat the dissociative process in this work. The frame transformation radius was set
at the value $r_0$ = 7 bohr radii, chosen here as the shortest possible distance beyond which the 
electron-cation interaction $V(R,r_0)$ can be neglected. 
Fig.~\ref{fig-gg_07} shows the comparison in the energy window
0--0.5~eV but the disagreement is very similar throughout the entire interval we have computed: 0--2~eV.

\begin{figure}[tbh]
\includegraphics[width=0.48\textwidth]{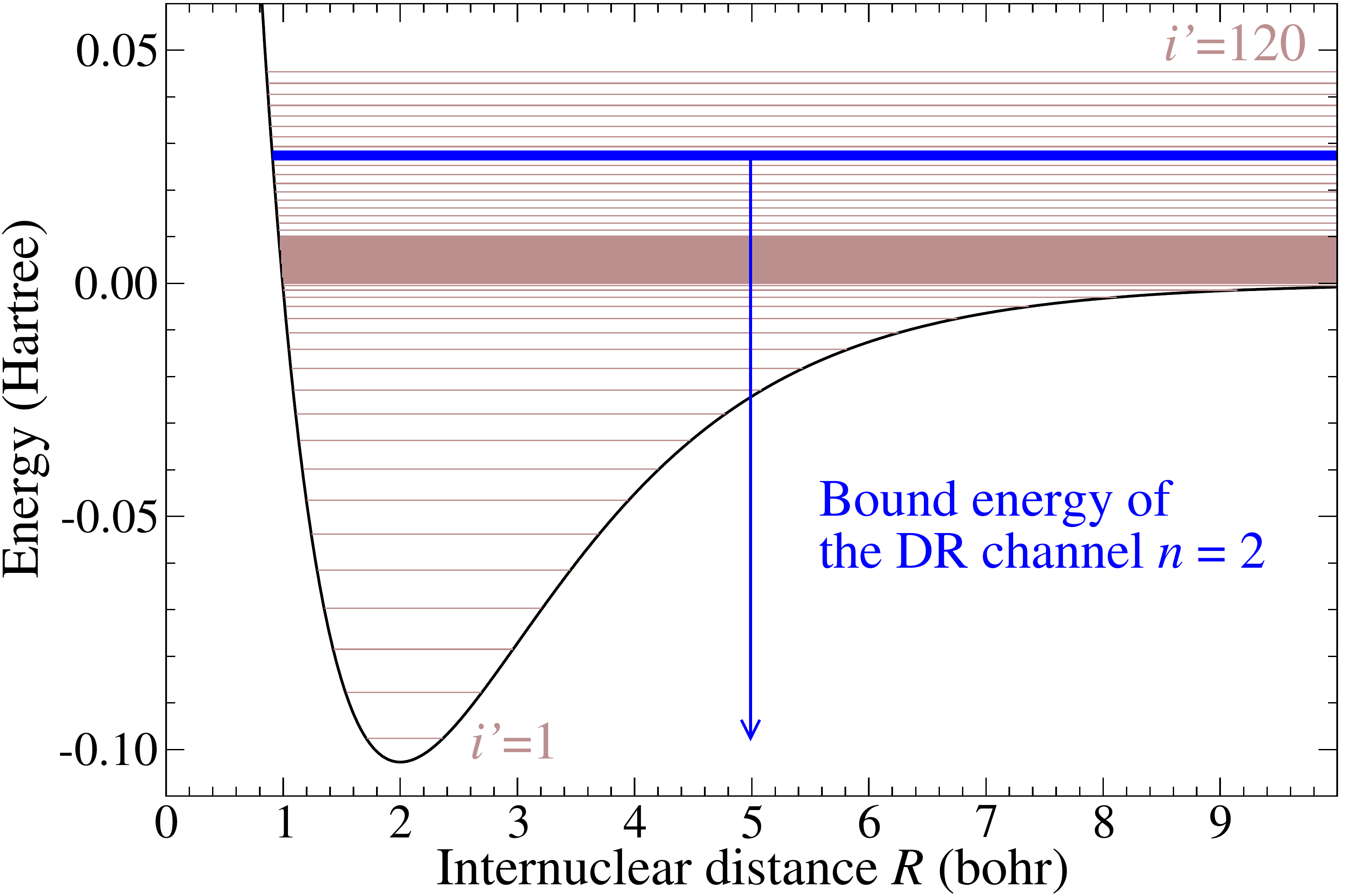}
\caption{\label{fig-pot}
The potential energy of the target cation is plotted versus the coordinate $R$ of the model
Hamiltonian.
Real parts of the vibrational energies of the
nuclear functions $\phi_{i'}(R)$ are displayed as horizontal lines. 
Length of the vertical arrow displays the binding energy (close to -1/8 Hartree) 
of one of the outgoing atomic fragments in its $n=2$ state.
The thick horizontal line is the kinetic energy of the nuclei in the $n$ = 2 DR channel for 
zero incident electron energy. 
}
\end{figure}

One of our previous publications \cite{Curik_HG_PRA_2018} indicated that the reason for the failure
of the energy-dependent FT theory applied to the DR, can be connected with inaccuracy
of the Born-Oppenheimer approximation inside the sphere confined by $r_0$~=~7~bohr.
Fig.~\ref{fig-pot} displays the real part of the energy levels of all the 120 target ion states included in the
present study, for which the Born-Oppenheimer approximation 
(\ref{eq-gg-psi}) and (\ref{eq-gg-F}) is assumed by the FT theory.
Fig.~\ref{fig-pot} also shows, as the thick horizontal line,
the energy of the nuclei after their dissociation into the $n$~=~2 channel, triggered by zero-energy incident 
electrons.
The high density of states at low positive energies is caused by the chosen bending angle 
$\theta = 40^{\circ}$, and by the large value of $R_0$
of the complex contour for generation of the ECS basis. Fig.~\ref{fig-pot} is helpful to indicate
the minimum number
of the ECS nuclear states that are needed by these calculations in order to describe
dissociation into $n$~=~2 state in the 0--0.5~eV incident collision energy window. 

\begin{figure}[tbh]
\includegraphics[width=0.48\textwidth]{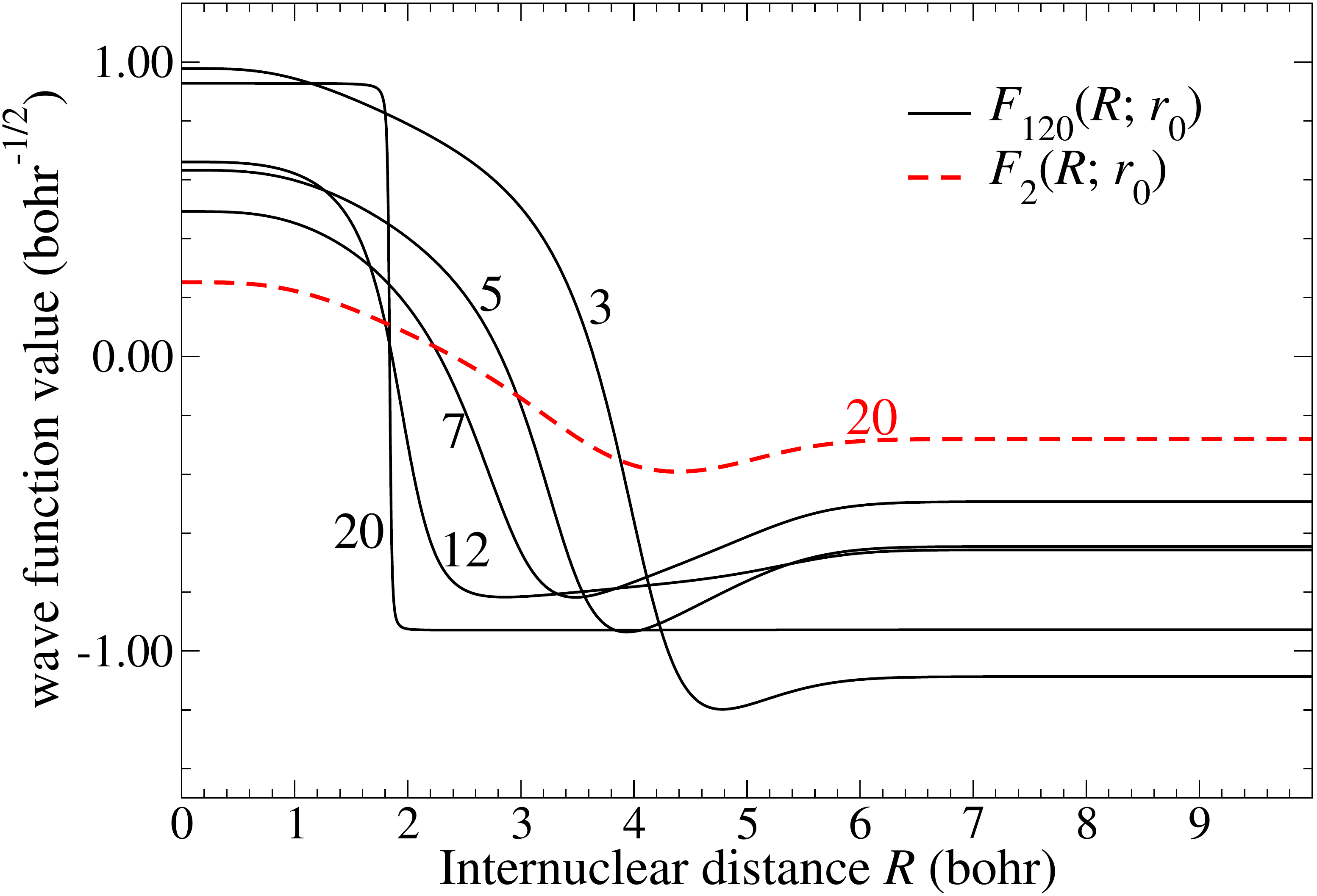}
\caption{\label{fig-wfr0}
Shape of the electronic wave function $F_{i'}(r_0;R)$ defined by Eq.~(\ref{eq-gg-F})
on the electronic surface $r_0$ for the following radii $r_0$ = 3, 5, 7, 12, 20 bohr. The
electronic energy corresponds to $i'$ = 120 for black curves and $i'$ = 2 for the dashed
curve.
}
\end{figure}

Validity of the BOA for the neutral complex in the present theory is connected with the size of 
$\partial F_{i'}(R; r)/\partial R$ for all the states included. Qualitatively, one can assess 
its accuracy by inspecting the
functions $F_{i'}(R; r_0)$ displayed in Fig.~\ref{fig-wfr0} for the highest $i'$~=~120 
state included and for different electronic radii $r_0$. As can be seen, as the electronic radius
$r_0$ increases, the wave function $F_{120}(R; r_0)$ changes from positive to negative value
over a smaller $R$-interval, generating a large magnitude of $\partial F_{i'}(R; r_0)/\partial R$ on the surface
$r_0$. The situation is less critical for lower states, e.g. the dashed line shows $F_2(R; r_0)$
at $r_0$~=~20~bohr. Thus good accuracy of the BOA inside even fairly large electronic sphere radii
can be expected for vibrational excitation studies. However, once the relevant dissociative
channels are included, the BOA leads to inaccurate results already at $r_0$~=~7 bohr
in the present study. Note that surface values of $F_{120}(R; r_0)$ for $r_0$~=~3, and 5 bohr
are computed inside the electron-cation interaction and the wave function cannot be considered
to reside in the asymptotic electronic region.

\section{\label{sec-bpft}Back-propagated frame transformation}

Breakdown of the Bohr-Oppenheimer approximation, shown for the dissociative nuclear
wave functions, leads to a question as to whether it is possible to decrease the frame transformation
radius $r_0$ to unphysically small values while still keeping all the information about
the electron-cation interaction. Such a procedure is indeed possible and has been designed here to 
consist of the three following steps:
\begin{enumerate}
\item
Determination of the energy-dependent quantum defect $\mu(R,\epsilon)$ at an appropriate electronic radius at which
the quantum defect is stable and converged, containing all the phaseshift 
relative to the Coulomb plus centrifugal potential which is acquired
in the electron-cation 
interaction. In the present study this value is approximately $r_0 \ge$~7~bohr. 
Knowledge  of the fixed-$R$ quantum defect allows to write the BOA solution $F(R;r)$ 
in Eq.~(\ref{eq-gg-F}) for $r > r_0$.
\item
Back-propagation of the electronic BOA solution $F(R;r)$ in the Coulomb field only to small distances $r_1$,
while ignoring the electron-molecule interaction $V(R,r)$, even though it is clearly 
non-negligible at the small distance $r_1$.
\item
The energy-dependent frame transformation of the back-propagated solutions
at $r_1$ then proceeds as is described in Section~\ref{sec-GGFT}. 
\end{enumerate}

\begin{figure}[bht]
\includegraphics[width=0.48\textwidth]{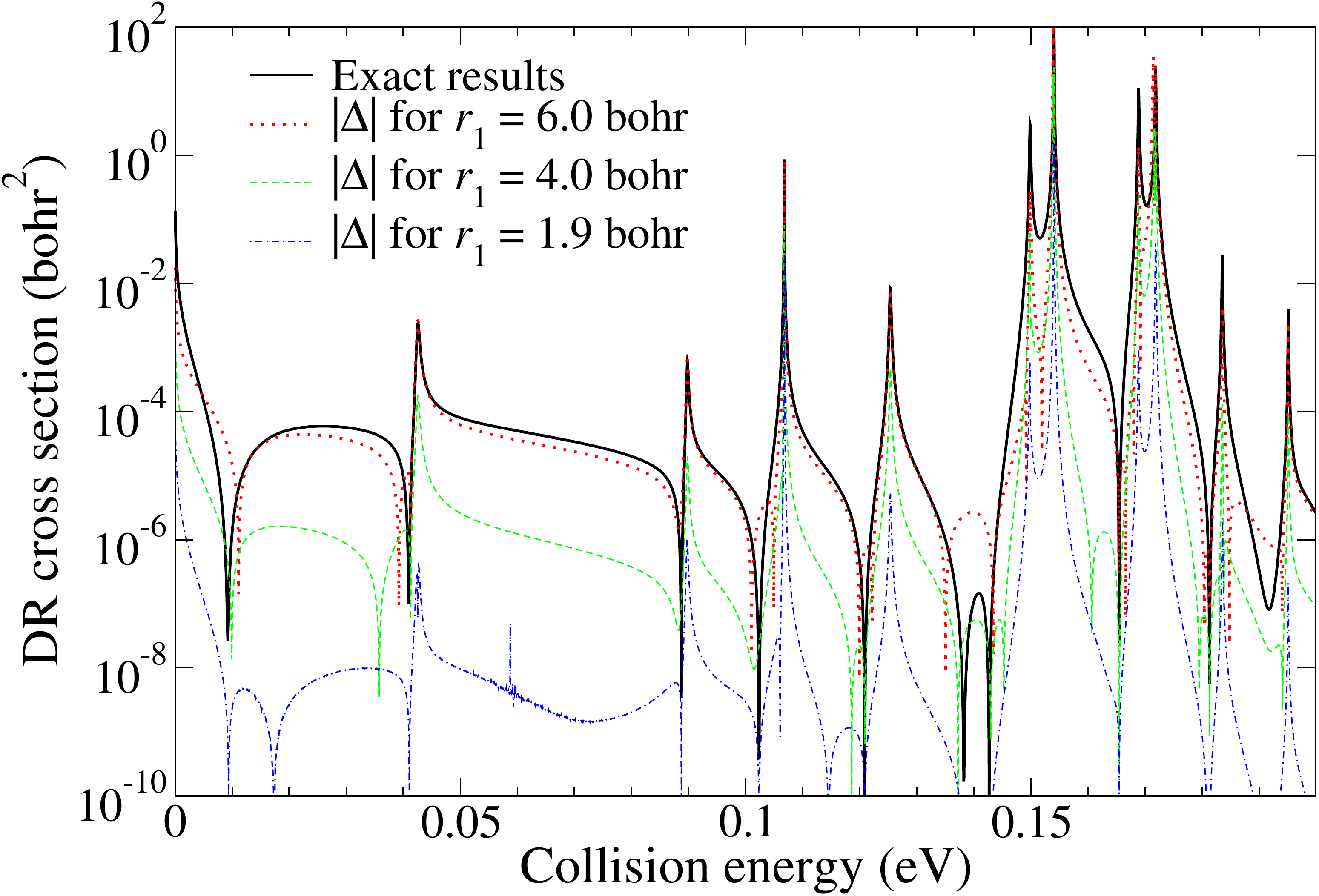}
\caption{\label{fig-csdr}
DR cross sections for different back-propagation distances. The thick line shows the absolute exact results.
Remaining data (denoted by the symbol $|\Delta|$) show absolute values of a difference between the back-propagated and exact results.
The back-propagation distances were $r_1$~=~6.0~bohr (dotted line), 4.0~bohr (dashed line), and
1.9~bohr (dot-dashed line).
}
\end{figure}

This proposed procedure is very simple to implement in practice. Once the full 
$\mu(R,\epsilon)$ is determined, the back-propagation of the electronic BOA functions
(\ref{eq-gg-F}) is implemented by a simple evaluation of the Coulomb functions $f_{i'}$ and
$g_{i'}$ at smaller different electronic radius $r_1$. Both steps 1. and 2., are executed simply by
replacing $r_0$ in Eqs.~(\ref{eq-gg-F})--(\ref{eq-gg-IJ}) with $r_1 < r_0$. In cases where
$\partial \mu(R,\epsilon)/\partial \epsilon > 0$ the radius $r_1$ can even be pushed to zero.
In the present case $\partial \mu(R,\epsilon)/\partial \epsilon$ is negative and there
is a bottom limit $r_1$~=~1.9 bohr below which the BOA wave function $F(R;r)$ becomes difficult to
normalize with Eq.~(\ref{eq-gg-norm}) as the term in brackets becomes negative for some $R$-values.
This minimum value of $r_1 \sim 1.9$ bohr also is reasonable, in view of the fact that the Coulomb plus 
centrifugal potential for $l=1$ reaches its minimum value at $r=2$ bohr.

In order to quantitatively test this ad-hoc procedure we have applied it to the 2D model problem. The Born-Oppenheimer
wave functions $F(R;r)$ obtained at $r = r_0 = 7.0$~bohr were back-propagated to three different distances
$r_1$~=~6.0, 4.0, 1.9 bohr radii.
Absolute values of the difference between the back-propagated and exact results (denoted as $|\Delta|$), 
together with the exact cross sections,
are shown in Fig.~\ref{fig-csdr}. The data demonstrate that the back-propagation step remarkably improves the
FT results. Back-propagation results for $r_1$~=~1.9 bohr are within 0.1\% of the exact cross
sections. For clarity the comparisons are presented over a narrower energy window 0--0.2~eV, but our presented
conclusions remain valid over the entire region examined: 0--2~eV.

\subsection{\label{ssec-Simple}Simplified version}

Simplified version of the back-propagation procedure is based on properties
of the Coulomb functions $f_{\epsilon}(r)$ and $g_{\epsilon}(r)$, which lose their energy
dependence as $r$ approaches zero value. In this limit it is reasonable to assume that 
Eq.~(\ref{eq-gg-IJ}) simplifies to
\begin{equation}
\label{eq-simpl1}
I_{i i'} = C_{i i'}\;, \quad\quad J_{i i'} = S_{i i'}\;,
\end{equation}
and thus $\underline{K} = \underline{S}\, \underline{C}^{-1}$.
The normalization factor $N(R,\epsilon)$ (\ref{eq-gg-norm}) approaches high values as the back-propagation
radius $r_1$ is pushed to the limit. This limit is $r_1 \rightarrow 0$ for 
$\partial \mu(R,\epsilon)/\partial \epsilon > 0$, or some small finite value for 
$\partial \mu(R,\epsilon)/\partial \epsilon < 0$.
The normalization factor
does not exactly cancel out in the 
$\underline{S}\, \underline{C}^{-1}$ product due to right-index dependence of its energy argument 
in Eqs.~(\ref{eq-gg-SC}). However, if the cancellation of $N(R,\epsilon)$ is assumed by
\begin{eqnarray}
\nonumber
C_{i i'} = \int dR\, \phi_i(R) \cos\pi\mu(R,\epsilon_{i'}) \phi_{i'}(R)\;,\\
\label{eq-gg-simpl2}
S_{i i'} = \int dR\, \phi_i(R) \sin\pi\mu(R,\epsilon_{i'}) \phi_{i'}(R)\;,
\end{eqnarray}
the evaluation of the normalization factor can be avoided. Moreover, the back-propagation radius $r_1$ does
not explicitly enter the simplified back-propagation procedure anymore. 
\begin{figure}[bht]
\includegraphics[width=0.48\textwidth]{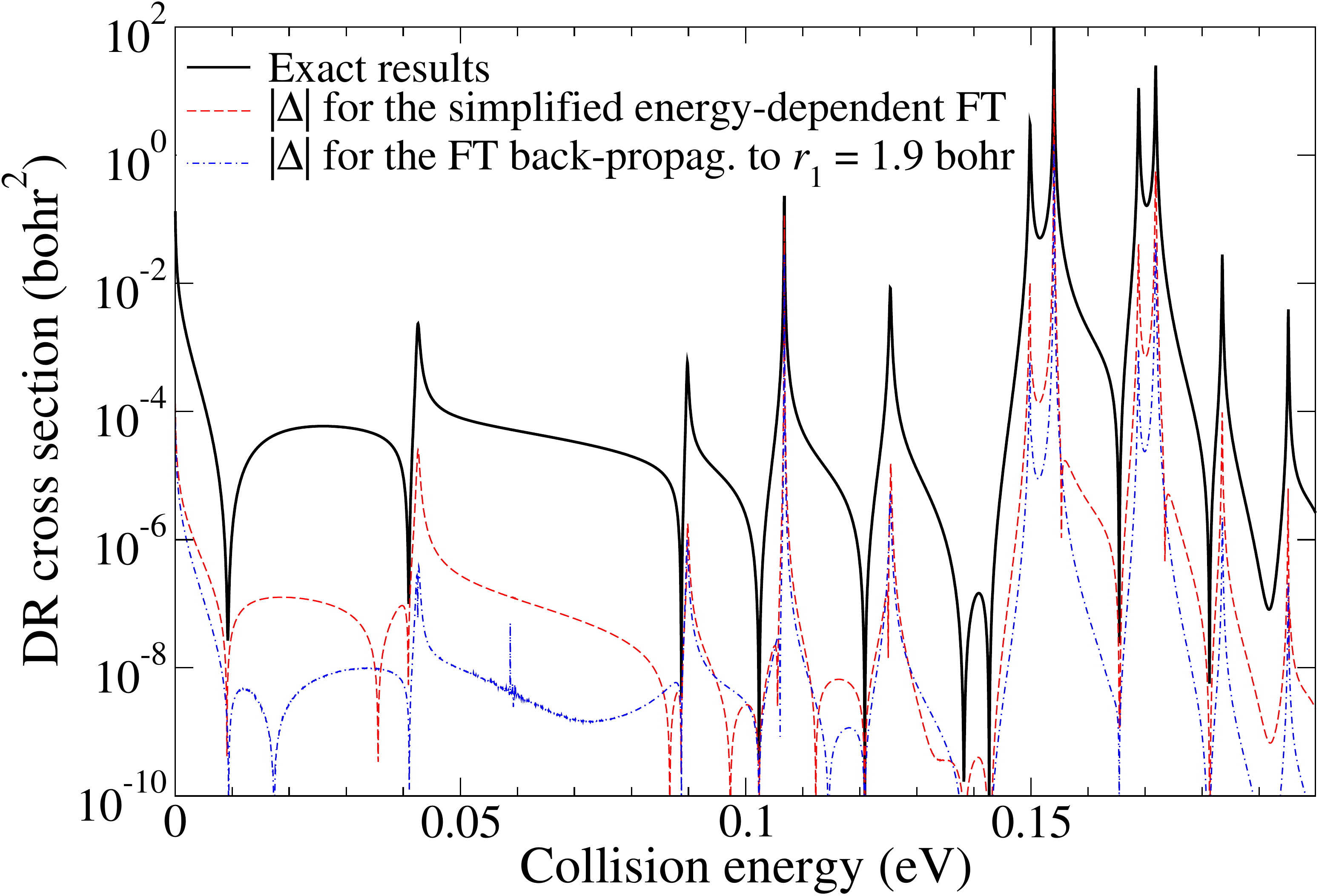}
\caption{\label{fig-csdrs}
DR cross sections for different back-propagation models. The thick curve shows the absolute exact results.
Data obtained by the back-propagated FT ($r_1$~=~1.9~bohr) are shown as a dot-dashed curve, while the
results of the simplified energy-dependent FT are displayed as a dashed curve. Both FT data sets display
absolute values of their difference from the exact results, as denoted by the symbol $|\Delta|$.
}
\end{figure}
Fig.~\ref{fig-csdrs} demonstrates that the simplification leads to a loss in accuracy of about one order
of magnitude, when compared to the back-propagated FT. However, regardless of its simplicity, the
simplified version, applied to the present model, yields results that are within 1\% accuracy from
the exact cross sections. It is also simpler to implement, in that it only requires knowledge of the 
body-frame quantum defect function and the vibrational wavefunctions, as the Coulomb functions 
${f,g}$ no longer appear in the integrals needed.

Finally, observe that the energy-independent frame transformation theory, widely
used in many of the recombination, vibrational excitation, and spectroscopic calculations,
can be viewed as a special case of the simplified version. This is because, 
the energy-independent FT theory neglects the energy dependence of the quantum defect, as well as
the energy dependence of the asymptotic Coulomb functions $f_{\epsilon}(r)$ and $g_{\epsilon}(r)$.

\subsection{\label{ssec-Simple}Vibrational excitation}

\begin{figure}[bht]
\includegraphics[width=0.48\textwidth]{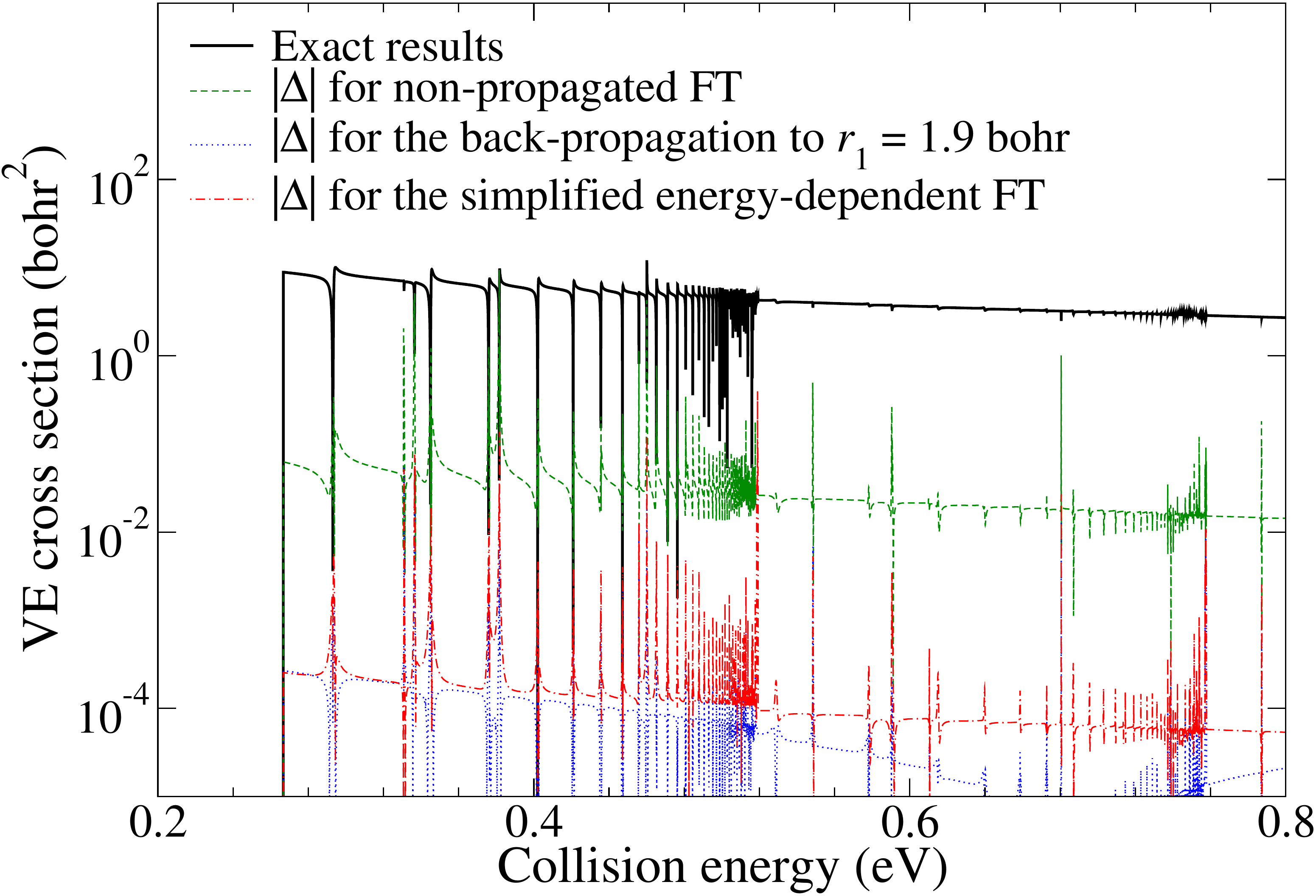}
\caption{\label{fig-csvib}
Vibrational excitation cross sections for transition $0 \rightarrow 1$. The thick curve shows the exact results.
Data obtained by the energy-dependent frame transformation without the back-propagation are shown by the
dashed line, the results of the back-propagated FT ($r_1$~=~1.9~bohr) are displayed by the dot-dashed
curve. Cross sections for the simplified back-propagated FT are shown by the dot-dashed curve.
All the three FT data sets display absolute values of their difference from the exact results, as denoted by the
symbol $|\Delta|$.
}
\end{figure}

In the previous section it was demonstrated that the back-propagation technique
leads to an improvement of the frame transformation theory for the dissociative recombination
process. The frame transformation theory can also be applied for another collisional
process that involves nuclear dynamics, namely vibrational excitation. 
The theoretical description of the vibrationally inelastic process also relies
on the Born-Oppenheimer approximation inside the inner region, however, as is qualitatively demonstrated
in Fig.~\ref{fig-wfr0}, to a lesser extent. This is clearly visible in Fig.~\ref{fig-csvib} displaying
that no back-propagation is necessary to obtain 1\% agreement between the energy-dependent FT theory
and the exact results.
However, also in the case of vibrational excitation process, the back-propagation procedure leads to further
improvement in the accuracy of the FT theory. Inaccuracies of the FT procedure are decreased 
by another two orders of magnitude, as is evident from Fig.~\ref{fig-csvib}. Note that the accuracies 
of the full back-propagation procedure and of its simplified version are comparable.

\section{\label{sec-concl}Conclusions}

The present study describes an extension of the energy-dependent frame transformation
theory by \citet{Gao_Greene_PRAR_1990} to dissociative recombination processes. The extension
is achieved by use of the complex outgoing-wave-type nuclear basis, implemented by exterior
complex scaling of the nuclear Hamiltonian. 
Direct application of the method is shown to yield inaccurate
results due to the limited validity of the Born-Oppenheimer approximation
for the dissociative processes. The demonstration is carried out on a simple but realistic 2D model system
tailored to describe the dissociative recombination of H$_2^+$ through the singlet ungerade channels,
which is an example of the indirect dissociative recombination process. Since the 2D model can be solved
exactly, to any desired numerical accuracy, its solutions provide an accurate benchmark to test
the frame transformation theory.

An additional procedure, based on the interaction-free back-propagation of the BOA solutions, 
is proposed to improve validity of the BOA inside the frame transformation radius. This ad-hoc
technique leads to a remarkable improvement of the computed DR cross sections, reproducing
the exact results within 0.1\% accuracy. A simplified version of the back-propagation 
procedure is also presented as a trade-off between accuracy and simplicity. Accuracy of the
simplified version is estimated to be within about 1\% of the exact results. 

The present study also qualitatively demonstrates that the vibrationally inelastic process is less sensitive
to failure of the Born-Oppenheimer approximation in the inner region, because its description
does not require such strongly closed channels as the ones that are necessary in the DR theory. Consequently, no
back-propagation is needed for vibrational excitation cross section calculations, provided 1\% accuracy 
is viewed as sufficient. However, application
of the back-propagation step improves the results, at least for the present model, by another two 
orders of magnitude in accuracy. In the case of vibrational excitation, the full back-propagation technique 
and its simplified version perform similarly in terms of accuracy.

The theory developed in this study, while tested here solely on a theoretical 2D model, 
has also been applied to the process of dissociative recombination in low energy collisions between electrons and HeH$^+$ cations. 
Application of the simplified version of the back-propagated FT is presented in Ref.~\cite{Curik_HG_PRL_2019}.

\begin{acknowledgments}
The work of CHG has been supported by the U.S. Department of Energy, Office of Science,
under Award No. DE-SC0010545; Basic Energy Sciences. R\v{C} acknowledges support of
the Czech Science Foundation (Grant No. GACR 18-02098S). Work of DH was supported
by the Czech Science Foundation (Grant No. GACR P203/17-26751Y).
\end{acknowledgments}

\bibliographystyle{apsrev}
\bibliography{DR}
\end{document}